\documentclass[lettersize,journal]{IEEEtran}
\usepackage{amsmath,amsfonts}
\usepackage{algorithmic}
\usepackage{array}
\usepackage[caption=false,font=normalsize,labelfont=sf,textfont=sf]{subfig}
\usepackage{textcomp}
\usepackage{stfloats}
\usepackage[noadjust]{cite}
\usepackage{url}
\usepackage{longtable}
\usepackage{verbatim}
\usepackage{graphicx}
\usepackage{float}
\usepackage{setspace}
\usepackage{subfig}
\usepackage{epstopdf}
\usepackage{booktabs}
\usepackage{multirow}
\usepackage{multicol}
\usepackage{threeparttable}
\usepackage{makecell}
\def\BibTeX{{\rm B\kern-.05em{\sc i\kern-.025em b}\kern-.08em
    T\kern-.1667em\lower.7ex\hbox{E}\kern-.125emX}}
\usepackage{balance}
\begin{document}
\title{Underwater Acoustic Signal Denoising Algorithms: A Survey of the State-of-the-art}
\author{
Ruobin Gao,~\IEEEmembership{Member,~IEEE,}
        % <-this % stops a space
Maohan Liang,
Heng Dong,~\IEEEmembership{Member,~IEEE,}
Xuewen Luo,~\IEEEmembership{Member,~IEEE,}
P. N. Suganthan,~\IEEEmembership{Fellow,~IEEE,}
\thanks{ \textit{Corresponding author: Maohan Liang}.}
\thanks{Ruobin Gao is with the School of Civil and Environmental Engineering, Nanyang Technological University, Singapore (e-mail: GAOR0009@e.ntu.edu.sg).}

\thanks{Maohan Liang is with the Department of Civil and Environmental Engineering, National University of Singapore, Singapore (e-mail: mhliang@nus.edu.sg).}

\thanks{Heng Dong is with the School of Electronics and Information Engineering, Harbin Institute of Technology, China (e-mail: dongheng@stu.hit.edu.cn).}

\thanks{Xuewen Luo is with the Communication Research Center, Harbin Institute of Technology, China, e-mail: luoxw@hit.edu.cn}

\thanks{P. N. Suganthan is with the KINDI Center for Computing Research, College of Engineering, Qatar University, Doha, Qatar (e-mail: p.n.suganthan@qu.edu.qa).}
\thanks{Manuscript created Jun, 2024; This work was developed by the IEEE Publication Technology Department. }}

\markboth{IEEE Transactions on SMC: Systems, Jun~2024}%
{How to Use the IEEEtran \LaTeX \ Templates}

\maketitle

\begin{abstract}
This paper comprehensively reviews recent advances in underwater acoustic signal denoising, an area critical for improving the reliability and clarity of underwater communication and monitoring systems. Despite significant progress in the field, the complex nature of underwater environments poses unique challenges that complicate the denoising process. We begin by outlining the fundamental challenges associated with underwater acoustic signal processing, including signal attenuation, noise variability, and the impact of environmental factors. The review then systematically categorizes and discusses various denoising algorithms, such as conventional, decomposition-based, and learning-based techniques, highlighting their applications, advantages, and limitations. Evaluation metrics and experimental datasets are also reviewed. The paper concludes with a list of open questions and recommendations for future research directions, emphasizing the need for developing more robust denoising techniques that can adapt to the dynamic underwater acoustic environment.
\end{abstract}


\section{Introduction}
\IEEEPARstart{U}{nderwater} Underwater acoustic data are crucial for various applications \cite{wu2019survey,zhang2023underwater}. The efficient and intelligent processing of these data is vital for enhancing state-of-the-art underwater technologies. While numerous technologies have been developed specifically for terrestrial and aerial environments, the unique characteristics of the underwater environment make acoustic signals particularly effective for capturing its conditions. However, severe noise interference presents significant challenges to receivers in underwater communications \cite{zhang2022modulation,singer2009signal,zhu2021convolutional}. The complex underwater settings, unpredictable transmission channels, and varying motion states significantly affect real-world underwater acoustic signals (UAS), potentially obscuring the inherent features of targets \cite{xie2024unraveling,jiang2022detection,liu2022review}. Consequently, developing advanced technologies for UAS denoising has become a critical and burgeoning research area in underwater scenarios.
\begin{table}[htbp]
\centering
  %\caption{Reconstruction error in the UAS denoising literature.}
  \label{tab:Abbreviation.}
  \resizebox{.5\textwidth}{!}{
\begin{tabular}{|ll|}
\hline
List of important abbreviations&\\
ADMM&Alternating direction method of multipliers\\

CEEMDAN&Complete ensemble EMD with adaptive selective noise \\
CNN&Convolutional neural network\\

DE&Dispersion entropy \\
DL&Deep learning\\
DWT& Discrete wavelet transform \\
EMD& Empirical mode decomposition \\
EEMD& Ensemble Empirical mode decomposition \\
EWT& Empirical wavelet transform \\
FT&Fourier transform\\
FFT&Fast Fourier transform\\
GAN&Generative adversarial networks\\
IBM&Ideal binary mask\\
ISTFT&Inverse short-time Fourier transform \\
IMF&Intrinsic mode function \\
LMS&Least mean squares\\
LWTD&Lift wavelet threshold\\
MAE&Mean absolute error\\
MFCC&Mel-frequency cepstral coefficient\\
MLP&Multilayer perceptron\\
MRA&Multi-resolution analysis\\
MSE&Mean Square Error\\
MSRU& Multiscale residual unit\\
PE&Permutation entropy\\
PSNR&Peak signal-to-noise ratio\\
RMSE&Root mean square error \\
RNN&Recurrent neural network\\
SDR&Signal-to-distortion ratio\\
SNR&Signal-to-noise ratio\\
SSNR&Segment signal-to-noise ratio\\
STFT&Short-time Fourier transform \\

VMD& Variational mode decomposition \\
UAS& Underwater acoustic signal \\
$x(t)$& Signal \\
$P_{signal}$& Power of the signal \\
$P_{noise}$& Power of the noise \\
$\phi_{j,k}(t)$& Scaling function\\
$\psi_{j,k}(t)$&Wavelet function \\
$\omega$&Normalized frequency \\
$\gamma$&Transitional band width parameter\\
$e(t)$&Envelope\\
$m(t)$& Mean of envelopes\\
$\lambda$&Threshold\\

$V$&Space\\
$K$&Number of modes\\
\hline
\end{tabular}}
\end{table}
\begin{figure}[htbp]
	\centering
	\includegraphics[width=\linewidth]{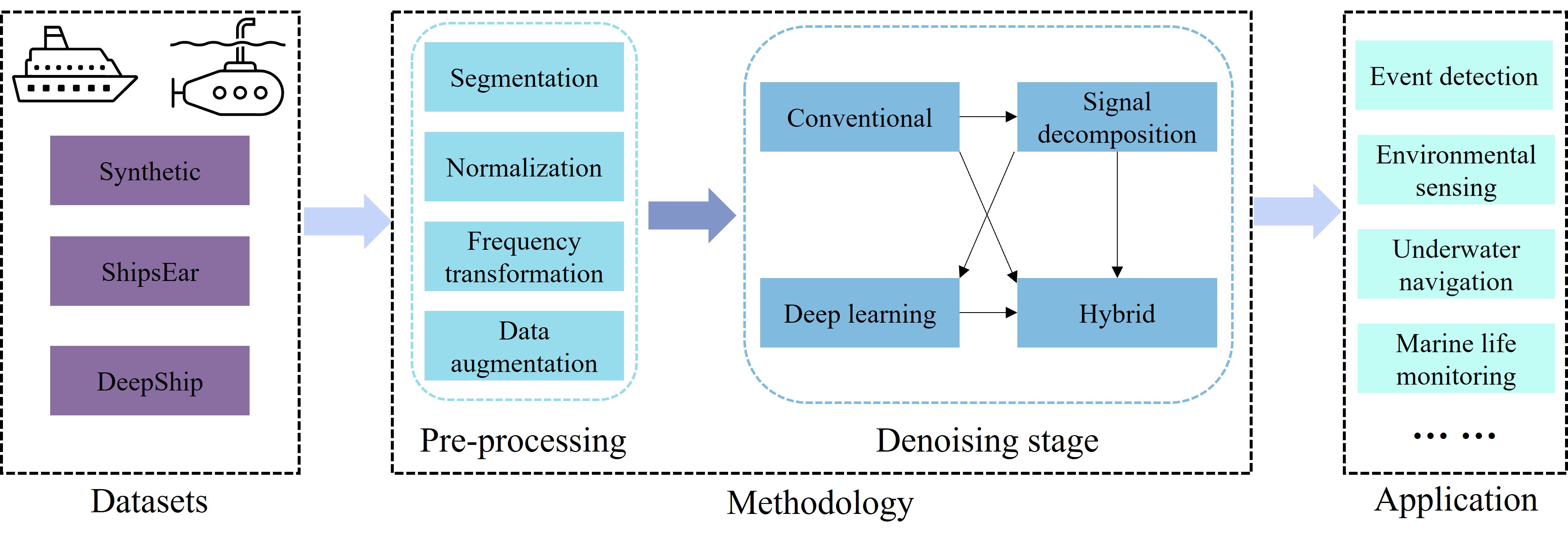}
	\caption{Framework of underwater acoustic signal denoising.}
	\label{fig:Pipeline_UAS_denoising}
\end{figure}
% handcrafted and signal processing
Since the UAS contains intensive noise, extracting noise-resistant features is essential for underwater recognition tasks \cite{jiang2021interpretable}. The UAS denoising can be categorized into four groups: Conventional approaches, decomposition-based framework, deep learning (DL) algorithm, and hybrid schemes. The overall framework of UAS denoising is shown in Figure \ref{fig:Pipeline_UAS_denoising}. Conventional frameworks employ handcrafted statistical measurements to obtain the feature set, which is utilized to train the learning-based recognition model \cite{jiang2021interpretable}. For instance, two recognition models based on neural networks are trained using eighty-eight features computed from the UAS \cite{jiang2021interpretable}. Although these handcrafted features are interpretable, they may not effectively capture the high-level abstractions in underwater data that are essential for tasks involving complex patterns. Moreover, defining suitable handcrafted features for specific tasks requires extensive domain knowledge and offers limited flexibility.\par
%decomposition-based framework
Inspired by the 'divide and conquer' strategy, the UAS denoising community has explored the decomposition of complex UAS into simpler components, which are then individually or collectively denoised. The family of signal decomposition algorithms in UAS denoising is extensive, including empirical mode decomposition (EMD) \cite{huang1998empirical}, variational mode decomposition (VMD) \cite{dragomiretskiy2013variational}, discrete wavelet transform (DWT) \cite{shensa1992discrete}, empirical wavelet transform (EWT) \cite{EWT}, and their advanced variants \cite{wu2009ensemble}. Following decomposition, criteria are established and applied to classify components into signal-dominated, noise-dominated, and pure noise groups. Specific processing or denoising techniques are then tailored to each category. Noise components are discarded, while signal components are preserved. The final step involves the aggregation of these processed components to produce a denoised UAS.

%deep learning 
In recent years, DL algorithms have succeeded across various fields due to their robust representation capabilities and minimal assumptions about the input data \cite{lecun2015deep}. The feature extraction prowess of DL algorithms has prompted researchers to explore their effectiveness in UAS denoising \cite{zhou2020denoising, sun2022underwater, tian2021deep, wang2014robust}. DL-based frameworks for UAS denoising typically utilize different neural network architectures to reconstruct clean signals from noisy inputs and maximize the signal-to-noise ratio (SNR). These models often employ an autoencoder architecture. The design of an efficient DL-based UAS denoising model depends on the choice of architectures and loss functions. The advantages of DL-based techniques include their consistency in denoising and applicability to subsequent tasks such as recognition or analysis based on UAS. Thus, the denoising process is task-oriented, aiming to ensure satisfactory performance across these applications.

%contribution
The underwater data mining and signal process community has been dedicated to imagery data \cite{jiang2020novel,zhou2023underwater,zhou2023underwater2}, but much less effort on acoustic data. Although there are some reviews about underwater sensing, they neglect a crucial role of UAS denoising \cite{lin2022deep,xu2023deep,domingos2022survey}. Meanwhile, a comprehensive review of the state-of-the-art UAS denoising research needs to be conducted. This article comprehensively reviews recent advances in UAS denoising and contributes to the literature from the following perspectives:
\begin{itemize}
    \item[1] Despite the extensive research on UAS denoising algorithms, a comprehensive review systematically summarizing and discussing these diverse approaches is absent. This deficiency poses a significant challenge for researchers and practitioners aiming to thoroughly understand the landscape of UAS denoising techniques.
    \item[2] We systematically analyze UAS denoising algorithms, from conventional signal processing methods to advanced DL algorithms. Furthermore, we introduce a taxonomy of UAS denoising techniques, marking the first instance of such classification in the literature. This taxonomy meticulously delineates the current landscape of UAS research, revealing insightful connections among each category.
    \item[3] We outline the prevailing challenges encountered in UAS denoising and explore potential solutions. Spanning from methodological intricacies to real-world applications, these challenges offer valuable insights and point towards promising avenues for future research in UAS denoising. 
    \item[4] We elucidate the diverse applications of UAS denoising techniques, underscoring their essential role in various underwater applications. This exploration holds significant interest for readers and practitioners alike, highlighting the critical importance of UAS denoising in underwater contexts.
\end{itemize}

%\section{Preliminaries}

\begin{figure*} 
    \centering
  \subfloat[Annual distribution of recent publications.\label{fig:Annual distribution of UAS denoising publications}]{%
       \includegraphics[width=0.4\linewidth]{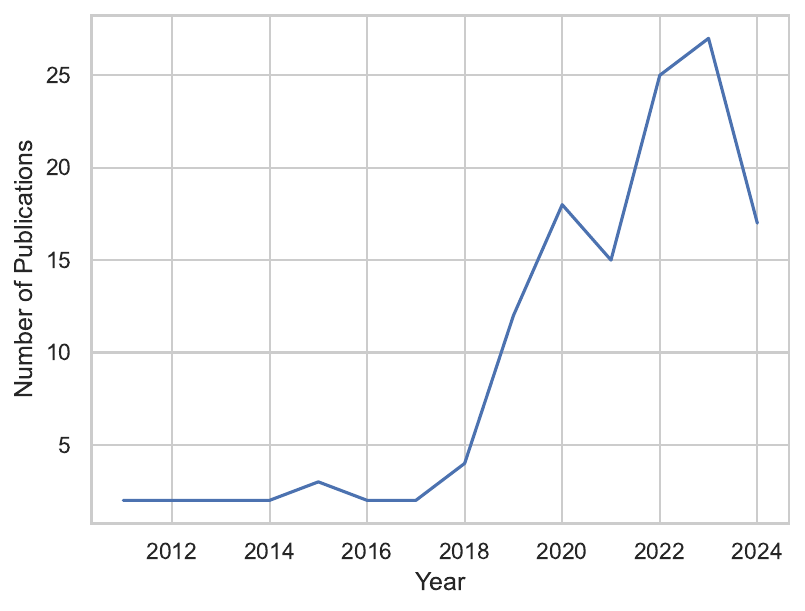}}
  %  \hfill
  %\hspace{2pt}
  \subfloat[Top ten authors by number of publications\label{FIG:Top ten Authors by Number of Publications}]{%
        \includegraphics[width=0.4\linewidth]{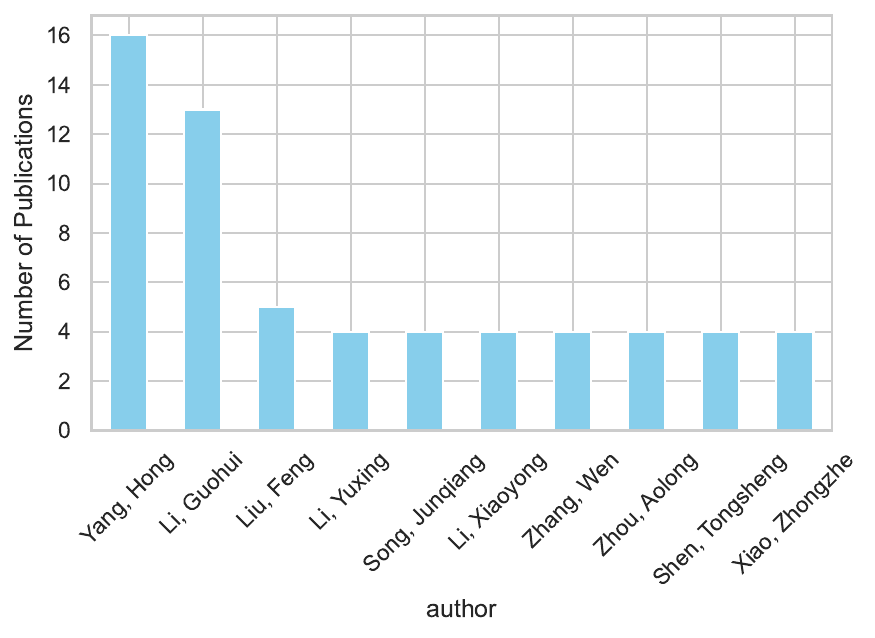}}
        \hspace{2pt}
     \subfloat[Top ten publication venues\label{FIG:Top ten publication venues}]{%
        \includegraphics[width=0.4\linewidth]{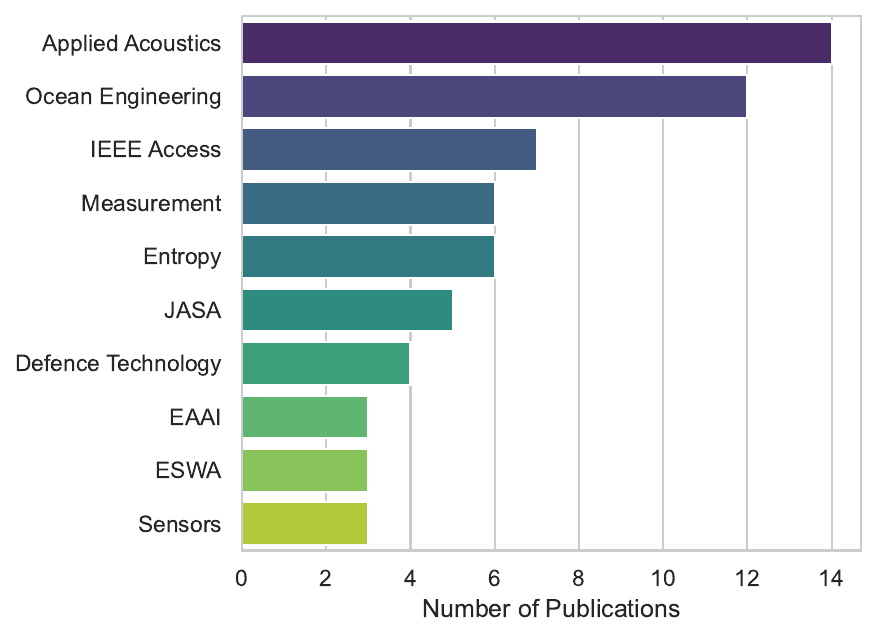}}
        \hspace{2pt}
         \subfloat[Network graph of co-Authorship\label{FIG:Network graph of co-Authorship}]{%
        \includegraphics[width=0.4\linewidth]{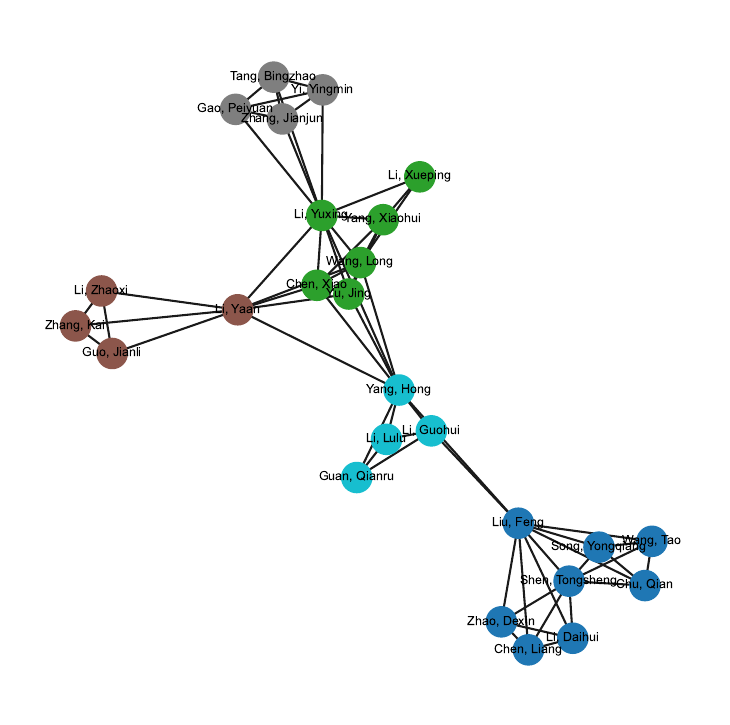}}
        \hspace{2pt}

  \caption{Bibliometric analysis of UAS denoising from 2011 to 2024.}
  \label{fig:BIB ANALYSIS} 
\end{figure*}

\section{Overview of UAS denoising research}

This section first conducts a bibliometric analysis of the reviewed UAS denoising literature. Then, we discuss the unique challenges in denoising UAS data.
\subsection{Bibliometric Analysis}\label{sec:bib}

This survey reviews research in UAS denoising techniques, predominantly published in academic journals or conferences related to ocean engineering, measurement, signal processing, and artificial intelligence (AI). Notable venues include \textit{IEEE Transactions on Instrumentation and Measurement}, \textit{Ocean Engineering}, \textit{IEEE Journal of Ocean Engineering}, \textit{Journal of Marine Science and Engineering}, \textit{Applied Acoustics}, \textit{Applied Ocean Research}, \textit{Measurement}, and \textit{The Journal of the Acoustical Society of America}. Figure~\ref{fig:BIB ANALYSIS} visualizes the scope of UAS denoising studies addressed in this survey. Figure~\ref{fig:Annual distribution of UAS denoising publications} illustrates a consistent upward trend in the number of studies within the UAS denoising field, despite a drop in publications in 2024, which only encompasses the first half of the year. Figure~\ref{FIG:Top ten Authors by Number of Publications} summarizes the top ten authors by the volume of their contributions to UAS denoising research. According to Figure~\ref{FIG:Top ten publication venues}, the journals \textit{Applied Acoustics} and \textit{Ocean Engineering} publish the most research on UAS denoising, given their specific focus on acoustics and ocean engineering. Recently, with advancements in AI, AI-related journals such as \textit{EAAI} and \textit{ESWA} have also shown increased receptivity to UAS denoising research. Finally, Figure~\ref{FIG:Network graph of co-Authorship} displays the network graph of co-authorship within the UAS denoising literature, highlighting five collaborative communities and the extent of their interactions.

\subsection{Why UAS denoising is challenging}
\label{sec:why uas denoising}
\begin{figure}[htbp]
	\centering
	\includegraphics[width=\linewidth]{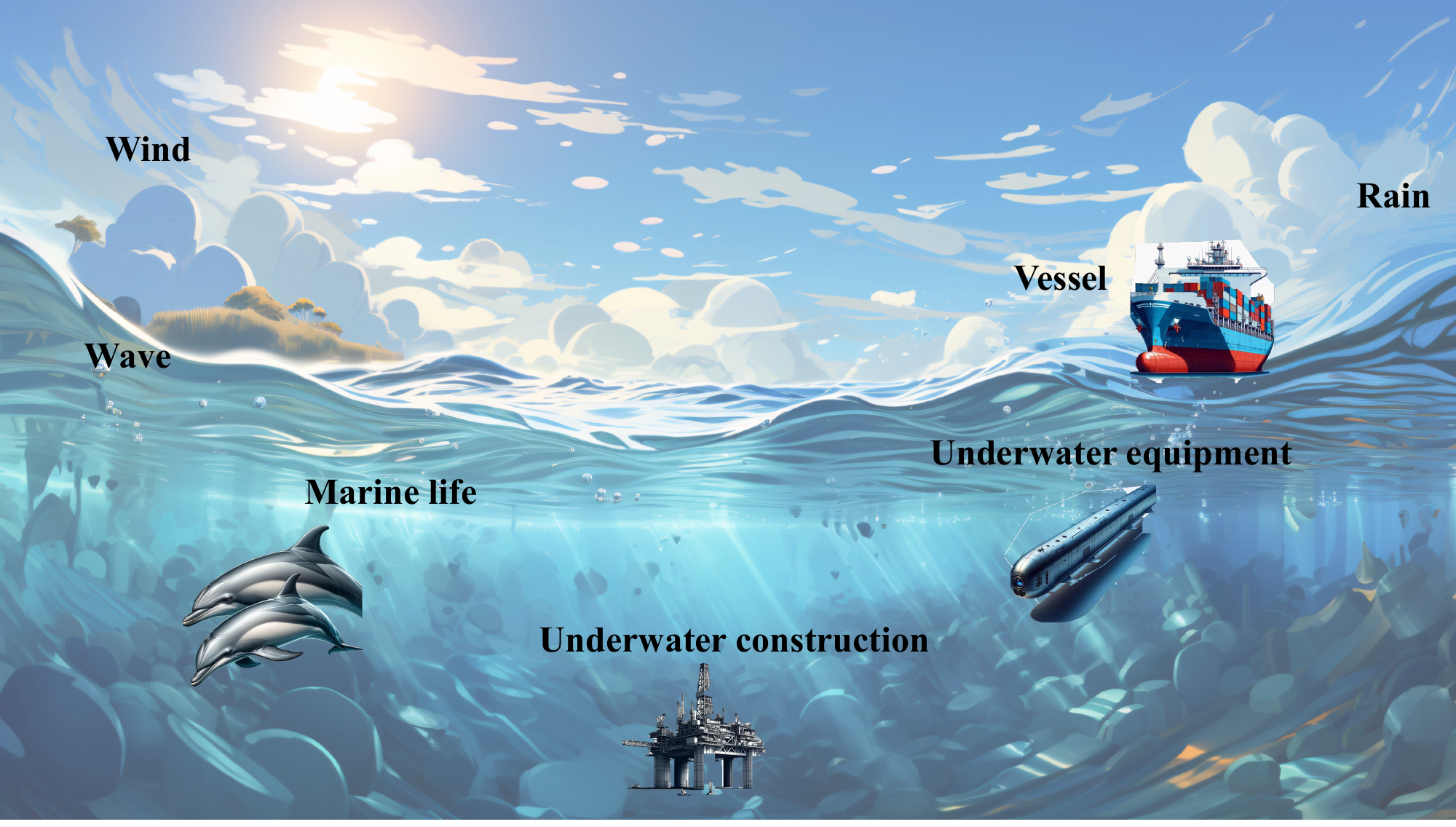}
	\caption{Underwater scenario illustrating complex noise sources.}
	\label{fig:UnderWaterNoiseSource}
\end{figure}
The low SNR of the UAS presents significant challenges; however, the complexities of the underwater environment introduce unique difficulties that distinguish UAS denoising from typical signal denoising tasks \cite{sendra2015underwater,singer2009signal}. Various noise sources exist in underwater environments, as shown in Figure \ref{fig:UnderWaterNoiseSource}. In underwater settings, the noise sources vary and include natural and anthropogenic elements. Natural sources such as marine life activity, wind-driven waves, and precipitation contribute significantly to the background noise. On the other hand, anthropogenic sources include ship traffic, industrial activities, sonar systems, and other man activities, all adding complexity to the noise environment \cite{duarte2021soundscape,butler2022habitat}. These diverse noise sources necessitate specialized approaches in UAS denoising to effectively separate the signal from the noise, ensuring clarity and accuracy in data interpretation.
\subsubsection{Complex sources}
In the underwater scenario, the noise sources contributing to the complexity of acoustic signals can be broadly categorized into natural, anthropogenic, and system-based sources, each adding layers of challenges to UAS denoising:

1. \textbf{Natural Noise Sources}:
\begin{itemize}
   \item  \textit{Biological Noise}: This includes sounds from marine life, such as whales, dolphins, and fish \cite{pieretti2020anthropogenic}. These biological entities often produce sounds for communication, navigation, and foraging, which can overlap in frequency and time with the signals of interest.
   \item  \textit{Geophysical Noise}: Phenomena such as wind, rain, and sea state contribute to background noise \cite{hilmo2020physical}. Turbulence caused by waves breaking on the surface and interactions between water and seabed during storms generates significant noise levels \cite{zhang2022review}.
   \item  \textit{Thermal Noise}: Caused by the random motion of water molecules, thermal noise is more prevalent in deeper and warmer waters and acts as a constant background noise across all frequencies \cite{chen2022routing}.
\end{itemize}
2. \textbf{Anthropogenic Noise Sources}:
\begin{itemize}
   \item \textit{Shipping Traffic}: Noise from commercial and recreational vessels is a dominant noise source in many oceanic environments. The sound from engines, propellers, and hull movement is pervasive at various frequencies and intensities \cite{cominelli2020vessel,jalkanen2022underwater}.
   \item \textit{Industrial Activities}: Underwater construction, oil drilling, and other marine operations involve heavy machinery that emits substantial acoustic signals \cite{chou2021international,halliday2020underwater}.
   \item \textit{Sonar and Naval Exercises}: Active sonar systems used by the military and some commercial ships emit powerful sound pulses that can interfere with and mask natural acoustic signals \cite{isojunno2020noise}.
\end{itemize}
3. \textbf{System-based Noise Sources}:
\begin{itemize}
   \item \textit{Instrument Noise}: Noise inherent to the recording devices, such as electronic noise from sensors and recording equipment, can affect the data quality \cite{cope2021multi}.
   \item  \textit{Data Transmission Noise}: In wireless underwater communication, signals can be corrupted by noise introduced during transmission, including reflections and refractions from the water's surface or the seabed.
\end{itemize}

Each noise source interacts differently with the underwater environment, making it challenging to isolate and remove unwanted noise from valuable data. Effective denoising thus requires a deep understanding of both the characteristics of these noises and the acoustic properties of the environment. Advanced signal processing techniques, adaptive filtering, and machine learning models are typically employed to enhance the clarity and reliability of the extracted signals in such complex scenarios.
\subsubsection{Energy imbalance}
In underwater environments, the fusion of multi-source signals frequently results in an imbalanced energy distribution within the captured acoustic data. This imbalance complicates the signal processing tasks, particularly the denoising of UAS. Factors such as varying signal intensities, overlapping frequency ranges, and sporadic or persistent noise sources further exacerbate the challenge. These complexities necessitate sophisticated denoising techniques that effectively distinguish between noise and actual acoustic signals of interest. Moreover, the dynamic nature of underwater environments, including changes in water density, temperature, and movement, adds additional layers of variability that denoising algorithms must account for. Consequently, improving the accuracy of UAS denoising involves addressing the imbalance in signal fusion and adapting to the inherently noisy and unpredictable underwater acoustic landscape.
\subsubsection{Disparate optimization objectives}
Since UAS denoising is usually the first stage for underwater recognition tasks, recognition models must be developed. Most literature treats denoising and recognition as two independent stages \cite{zhou2023novel}. When designing UAS denoising algorithms, researchers may not consider the requirements of recognition tasks. The denoising stage is unsupervised, and recognition labels are unavailable. The objectives of developing denoising and recognition models are different, challenges in unifying these two stages. 
%\section{Problem formulation}
%Suppose that the collected noisy UAS is
%\begin{equation}
 %   y(t) = x(t)+n(t), 
%\end{equation}
%where $n(t)$ is the noise, $x(t)$ and $y(t)$ are clean signal and noisy signal. The goal of UAS denoising algorithm is to eliminate $n(t)$ and retain pure signal $x(t)$. 
%\section{Denoising methods}
\section{Conventional methods}
Conventional UAS denoising is usually based on hand-craft features \cite{jiang2020multi} and linear filtering \cite{al2017improved}. UAS is split into various barks denoised by wavelet thresholding algorithms \cite{wang2014robust}.
Frame-Based Time-Scale Filters method is proposed to improve the standard wavelet soft-thresholding in reducing distortions in the joint time-frequency space \cite{ou2011frame}. A two-stage denoising framework consisting of adaptive window median filter and wavelet threshold optimization is designed to eliminate Gaussian and non-Gaussian noise, respectively \cite{wang2020novel}. This article focuses on the most recent advancement in UAS denoising. For conventional signal processing and thresholding techniques, we suggest referring to these survey studies \cite{middleton1987channel,wenz1972review,awan2019underwater}.

\section{Signal decomposition}
\begin{figure}[htbp]
	\centering
	\includegraphics[width=\linewidth]{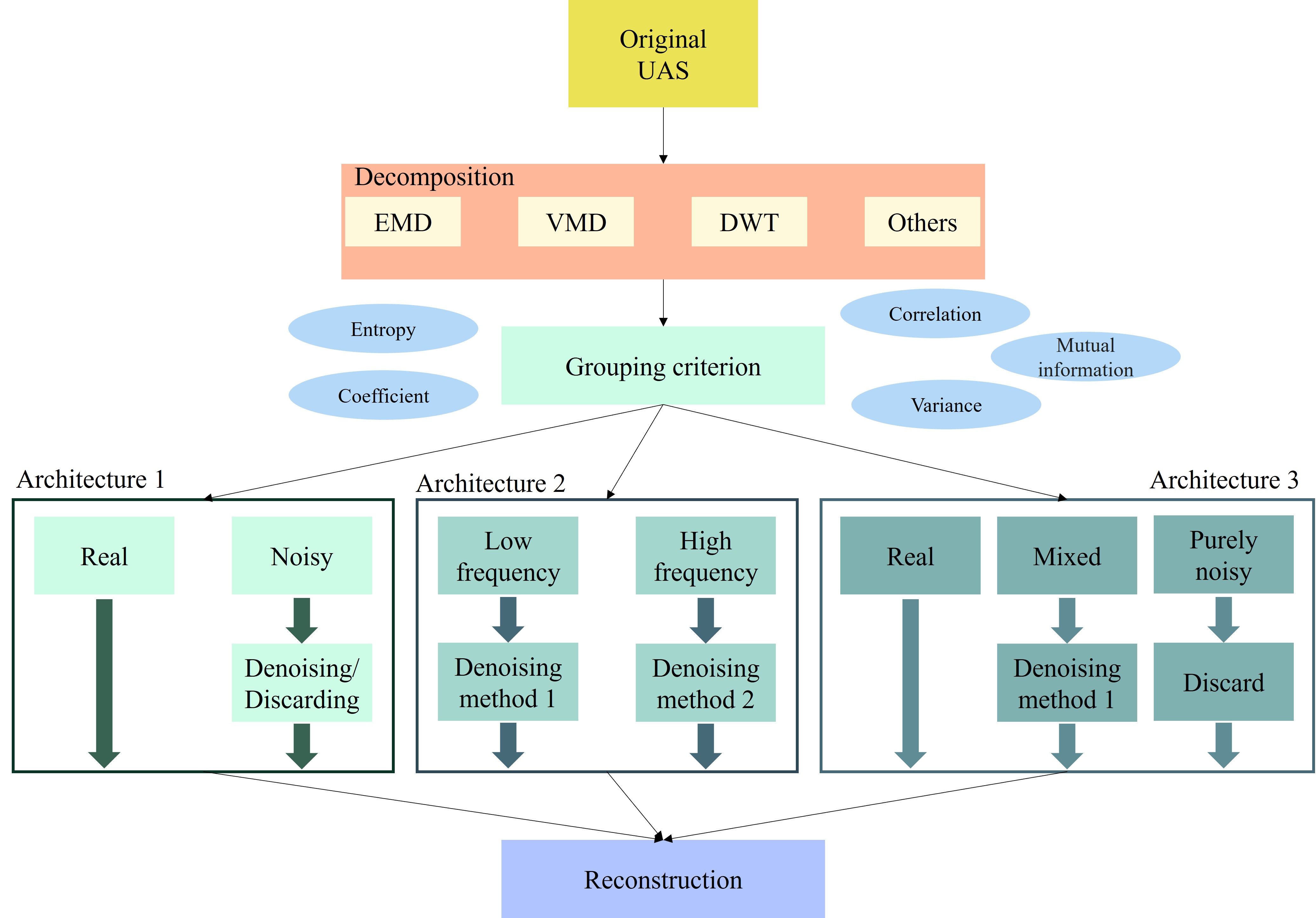}
	\caption{Framework of decomposition-based UAS denoising.}
	\label{fig:Pipeline_decomposition_denoising}
\end{figure}

Signal decomposition techniques can decompose complex signals into various components or modes, carrying information of different frequencies. Individual modes are easier to analyze, process, and denoise \cite{gao2023online}. The general framework of decomposition-based denoising methods is shown in Fig. After obtaining modes with the help of decomposition, suitable denoising algorithms are applied to these modes. Finally, denoised modes are aggregated to reconstruct the input signal. The overall framework of decomposition-based UAS denoising is visualized in Fig. \ref{fig:Pipeline_decomposition_denoising}.


\begin{table}[htbp]
  \centering
  \caption{Summary of representative decomposition-based UAS denoising studies}
  \resizebox{.5\textwidth}{!}{
    \begin{tabular}{ccccc}
    \toprule
    \multicolumn{1}{c}{Literature} & \multicolumn{1}{c}{Signal decomposition} & \multicolumn{1}{c}{Denoising} & \multicolumn{1}{c}{Grouping} & \multicolumn{1}{c}{Hyper-parameters} \\
    \midrule
    \cite{5972413}     & DWT     & Wavelet Thresholding & NA   & Manually \\
    \cite{li2024new}     & EMD     & Evolutionary filtering   & Entropy  & Sand cat swarm  \\

    \cite{li2024noise}     & EMD     & Wavelet Thresholding   & Entropy  & Manually \\
    
    \cite{li2024ultrasound}     & VMD     & Wavelet Thresholding   & Correlation   & Grey Wolf Optimisation \\
    \cite{liu2024application}     & DWT, VMD     &Wavelet Thresholding   & NA    & EMD's level \\
\cite{li2023research}     & VMD     &Discard noisy IMFs   & Entropy    & VMD's level \\

\cite{yang2023dual}     & VMD     &Grey relation analysis and thresholding  & Frequency    &  Manually\\

   \cite{li2022research}     & EMD     &Interval thresholding  & Entropy    &  Manually\\
\cite{yang2021denoising,yang2020denoising}     & VMD     &Wavelet thresholding  & Entropy    &  VMD's level\\
    
    \cite{li2020novel}     & EMD     &LMS filtering    & Mean square variance  &  Manually\\

    \cite{li2019denoising}     & EMD     &Interval thresholding  &Entropy  &  Manually\\
 \cite{hu2019denoising}     & VMD     &Wavelet thresholding  &Correlation and frequency &  Manually\\

    \cite{li2021novel}     & EWT    &Discard noisy components  &Energy intensity &  Manually\\
    \bottomrule
    \end{tabular}%
    }
  \label{tab:addlabel}%
\end{table}%

\subsection{Wavelet transform}
\subsubsection{Theoretical development}
Fourier transform (FT) has historically been the method of choice for spectral analysis until the advent of wavelet transform. The Fourier transform's limitations, particularly its inefficiencies in local time-frequency representation and its poor performance with non-stationary signals, have led to its replacement by wavelet transform. This newer method has subsequently demonstrated significant success in the analysis of time series data. DWT is calculated as Equation \ref{eq:DWT},
\begin{equation}
\label{eq:DWT}
\mathit{f(j,k)=<x(t),\psi_{j,k} (t)>=\int x(t)\psi_{j,k}^{*} (t){\mathrm{d} t}},
\end{equation}
where $\psi_{j,k} (t)=2^{\frac{j}{2}}\psi(2^{j}t-k),j,k\in \mathbb{Z}$ is the wavelet function in DWT. In a practical forecasting problem, signal $x(t)$ and $\psi_{j,k}(t)$ are both discrete as $t$ is the discrete time index. In reality, finite-length times series $x(t)\in L^{2}(R)$ are all applicable to DWT.

In 1988, Daubechies first introduced the construction of a finite-support orthogonal wavelet named the db wavelet family. For a specific wavelet, there is a pair of scaling function $\phi_{j,k}(t)$ and wavelet function $\psi_{j,k}(t)$ for scale $j$. 

The most important property for a scaling function and wavelet function to satisfy the multi-resolution analysis (MRA) is the dilation equation in Equation \ref{eq:phi dilation} and \ref{eq:psi dilation} based on which MRA is built. It indicates that the coarser basis $\phi (t)$ with larger support is a weighted sum of the finer basis $\phi (2t-k)$ with shorter support and $\{h_{k}, k\in\mathrm{Z}\}$ is the weight.
\begin{equation}
\label{eq:phi dilation}
\phi(t)=\sqrt{2}\sum_{k\in \mathrm{Z}}h_{k}\phi(2t-k)
\end{equation}
\begin{equation}
\label{eq:psi dilation}
\psi(t)=\sqrt{2}\sum_{k\in \mathrm{Z}}g_{k}\phi(2t-k)
\end{equation}
As $\left \{ \phi_{j,k}(t),k\in\mathrm{Z} \right \}$ spans space of scale $j$ ($V_{j}$), any function $f_{j}(t)$ in $V_{j}$ can be written as a linear combination of the orthogonal base 
$\left \{ \phi_{j,k}(t),k\in\mathrm{Z} \right \}$ 
as
\begin{equation}
\label{eq:wavebasis express}
f_{j}(t)=\sum_{k=-\infty }^{\infty} c_{j}[k]\phi_{j,k},
\end{equation}
where $c_{j}[k]$ is the coefficient of a corresponding basis $\phi_{j,k}$. Based on this representation and Equation \ref{eq:phi dilation}, \ref{eq:psi dilation}, the nestedness between space $\left \{ V_{j},j\in\mathrm{Z} \right \}$ can be derived as Equation \ref{eq:space relationship}.
\begin{equation}
\label{eq:space relationship}
\mathrm{V}_{j}\subset \mathrm{V}_{j+1}\subset \cdots \mathrm{V}_{J},\space \mathrm{j}<\mathrm{J}
\end{equation}

Moreover, the difference(residual) space of two adjacent spaces written as ${W}_{j}={V}_{j+1}-{V}_{j}$ is spanned by wavelet functions $\psi_{j,k}, (k\in \mathrm{Z})$ due to the orthogonality between $\phi_{j,k}$ and $\psi_{j,k}$ we can effortlessly know that ${W}_{j}\perp{V}_{j}$.

The scale \( j \) of \( V_j \) is contingent upon the selected wavelet, with different wavelet functions (bases) yielding distinct spaces. A fundamental principle of MRA dictates that the scaling function \( \phi_{j,k} \) exhibits a diminishing support length as \( j \) increases, which enhances its resolution and vice versa. Employing MRA in time series analysis provides an effective means to decompose a high-resolution signal into its constituent components, such as the rough trend and various cyclic frequencies, by utilizing different wavelet bases (scaling functions) at various scales.

\subsubsection{WT-based UAS denoising}
Classical wavelet threshold denoising techniques effectively suppress noise by leveraging thresholds derived from wavelet coefficients, thereby retaining stronger signals \cite{5972413,gur2007autocorrelation,ganapathi2016noise}. Thresholding within this context can be classified into hard, soft, and hybrid categories \cite{5972413}. However, classical wavelet thresholding algorithms struggle with non-Gaussian, non-linear, and non-stationary noise types. Moreover, selecting an appropriate threshold remains a significant challenge. To overcome these issues, some researchers have proposed using the posterior probability distribution of wavelet coefficients obtained from the DWT as the threshold to eliminate non-dominant coefficients \cite{liu2024application}. Additionally, integrating the lifting wavelet transform with soft thresholding has been investigated as a strategy to mitigate the shortcomings of the first-generation wavelet transform \cite{yang2020denoising}.

\subsection{Empirical wavelet transform}
\subsubsection{Theoretical development} 

The EWT represents an automated approach in signal processing, underpinned by robust theoretical foundations for decomposing non-stationary time series data \cite{gilles2013empirical}. Contrasting with DWT and EMD \cite{flandrin2004empirical}, EWT conducts a meticulous analysis of time series in the Fourier domain subsequent to the application of a Fast Fourier Transform (FFT). This technique involves the segmentation of the spectrum through data-driven band-pass filtering.

In the EWT, limited freedom is provided for selecting wavelets. The algorithm employs Littlewood-Paley and Meyer's wavelets because of the analytic accessibility of the Fourier domain's closed-form formulations \cite{spencer1994ten}. We represent the normalized frequency as $\omega \in [0,\pi]$. We utilize $\omega_{n}$ to represent the limits between the segments that are obtained from the Fourier support $[0,\pi]$. These band-pass filters' formulations are denoted using Equations \ref{eq:EWT_1} and \ref{eq:EWT_2}
	\begin{equation}
 \small
		\centering
		\label{eq:EWT_1}
		\hat{\phi}_{n}(\omega) = 
		\begin{cases}
			1& \makebox[1pt][r]{\text{if $\left|\omega\right| \leq (1-\gamma)\omega_{n}$}}\\
			
			\cos\left[\frac{\pi}{2}\beta\left(\frac{1}{2\gamma\omega_{n}}\left(\left|\omega\right|- (1-\gamma)\omega_{n}|\right)\right)\right]
			& \\
			\quad&\makebox[60pt][r]{\text{if $(1-\gamma)\omega_{n} \leq \left|\omega \right| \leq (1+\gamma)\omega_{n}$ }}\\
			
			0  & \text{otherwise,}\\
		\end{cases}
	\end{equation} 
	
	\begin{equation}
 \small
		\centering
		\label{eq:EWT_2}
		\hat{\psi}_{n}(\omega) = 
		\begin{cases}
			
			1& \makebox[12pt][r]{\text{if $(1+\gamma)\omega_{n} \leq \left|\omega \right| \leq (1-\gamma)\omega_{n+1}$ }}\\
			
			\cos\left[\frac{\pi}{2}\zeta\left(\frac{1}{2\gamma\omega_{n+1}}\left(\left|\omega\right|- (1-\gamma)\omega_{n+1}|\right)\right)\right]
			& \\
			\quad	&\makebox[22pt][r]{\text{if $(1-\gamma)\omega_{n+1} \leq \left|\omega \right| \leq (1+\gamma)\omega_{n+1}$ }}\\
			
			\sin\left[\frac{\pi}{2}\zeta\left(\frac{1}{2\gamma\omega_{n}}\left(\left|\omega\right|- (1-\gamma)\omega_{n}|\right)\right)\right]
			& \\
			\quad&\makebox[0.001pt][r]{\text{if $(1-\gamma)\omega_{n} \leq \left|\omega \right| \leq (1+\gamma)\omega_{n}$}}\\
			
			0  & \makebox[2pt][r]{\text{otherwise,}}\\
			
		\end{cases}
	\end{equation}
	
	with a transitional band width parameter $\gamma$ satisfying $\gamma \le \min_{n} \frac{\omega_{n+1}-\omega_{n}}{\omega_{n+1}+\omega_{n}}$. The most common function $\zeta(x)$ in Equations \ref{eq:EWT_1} and \ref{eq:EWT_2} are presented in Equation \ref{eq:Beta}. 
	\begin{equation}
		\label{eq:Beta}
		\centering
		\beta(x)=x^{4}(35-84x+70x^2-20x^3)
	\end{equation}
	
	This empowers the formulated empirical scaling and wavelet function
	$\{\hat{\phi}_{1}(\omega),\{\hat{\psi}_{n}(\omega)\}_{n=1}^{N} \}$ to be a tight frame of $L^{2}(\mathbb{R})$.	It can be observed that $\{\hat{\phi}_{1}(\omega),\{\hat{\psi}_{n}(\omega)\}_{n=1}^{N} \}$ are used as band-pass filters centered at assorted center frequencies. 
\subsubsection{EWT-based UAS denoising}
Although EWT achieved tremendous success in other sequence tasks, the UAS community has not researched its denoising ability \cite{gao2023online}. The EWT is employed to decompose the UAS into several sub-series and utilize the sub-series of highest energy to train a recognition model \cite{li2021novel}.

\subsection{Empirical mode decomposition}
\subsubsection{Theoretical development}
The EMD is a wholly data-driven approach for decomposing time-domain signals into distinct oscillatory modes and a residual component \cite{huang1998empirical}. Each mode, defined as an Intrinsic Mode Function (IMF), must satisfy two specific criteria to be classified as such.

1. The number of extremums in the oscillation and the number of zero crossings must equal or differ by at most one.

2. The mean of the envelopes defined by the local maxima and the local minima shall equal zero.

The signal decomposed by EMD can be expressed as the sum of a finite number of IMFs and a residual value. 
\begin{equation}
\label{emd_eq_q}
x(t)=\sum_{m=1}^k {IMF}_m(t)+r_k(t),
\end{equation}
where $k$ is the IMF number and $r_k(t)$ is the final residual value.

The set of IMFs constitutes a complete, adaptive, and nearly orthogonal basis for the original
signal. The algorithm for the iterative process of EMD is as follows: 
\begin{enumerate}
\item Find all the minima and maxima in $x(t)$.
\item Perform cubic spline interpolation of minima to obtain the lower envelope $e_m(t)$ and that of maxima for upper envelope $e_l(t)$.
\item Find the mean of the two envelopes using $m(t)=\frac{e_m(t)+e_l(t)}{2}$.
\item Subtract the mean from the signal as $d(t)=x(t)-m(t)$.
\item Applying the abovementioned factors, check whether $d(t)$ is an IMF.
\item If $d(t)$ is not an IMF, iterate from step (2) to (5) considering input as $d(t)$ to find the IMF.
\item If $d(t)$ is an IMF, find the residue $r(t) = x(t)-d(t)$.
\item If $r(t)$ has greater than two extrema, i.e., one maximum and one minimum and a single zero crossing the stopping criterion not satisfied, iterate from step (2) to (5), considering $r(t)$ as input to find the subsequent IMF.
\item If $r(t)$ has less than or equal to 2 extrema, i.e., one maximum, one minimum, and a single zero-crossing, the stopping criterion is satisfied, $r(t)$ is the final residue, and the EMD process is complete.
\end{enumerate}

The EMD method preserves the signal in the time domain. Each IMF encapsulates information on the variations in amplitude and frequency of the original signal over time. IMFs consist of a single or a narrow band of frequencies with no overlap. Furthermore, these functions or signals are orthogonal to the original signal \cite{zao2014speech}.
\subsubsection{EMD-based UAS denoising}
EMD and its variants have achieved tremendous success in UAS denoising \cite{kannan2015acoustic,wang2019underwater,li2024ultrasound}. The application of the EMD technique offers a novel approach to detecting and classifying marine mammal vocalizations in underwater acoustics, which traditionally requires extensive manual analysis by skilled acousticians. This method efficiently identifies and labels sound sources in a recording without prior knowledge or extensive pre-processing, streamlining the task through minimal post-processing quality control \cite{seger2018empirical}. The non-stationarity of each decomposition mode is utilized to select noise components obtained by the ensemble EMD (EEMD) \cite{caldeira2023eemd}.

The CEEMD is employed to denoise the original signal first, and a bidirectional denoising autoencoder is developed to learn robust representations \cite{dong2022bidirectional}. The complete ensemble empirical mode decomposition with adaptive selective noise (CEEMDAN) is adopted to decompose UAS into IMFs, and IMF with the minimum difference between the energy distribution ratio and average energy distribution ratio is selected \cite{yang2022novel}. While decomposing the acoustic target signal, the correlation coefficient between each IMF and the original signal is utilized as a threshold to determine signal-dominated IMFs. In addition to utilizing threshold to drop out noisy IMFs, the literature also employs denoising algorithms to denoise noisy IMFs \cite{li2020novel}. A criterion determining noisy IMFs is designed, and then noisy IMFs are denoised. Different criterion is proposed in the literature, such as minimum mean square variance \cite{li2020novel}, energy concentration property \cite{chengkun2024denoise}. For denoising IMFs, researchers have tried on least mean square filter \cite{li2020novel}.

Modified uniform EMD is employed to decompose the input into IMFs, and a double threshold is obtained according to hierarchical amplitude-aware permutation entropy (PE). The threshold assists in dividing IMFs into clean, mixed, and noisy IMFs. Since mixed IMFs contain noisy information, an evolutionary improved wavelet threshold denoising method denoises mixed IMFs \cite{li2024new}.

Recently, secondary decomposition outperforms one-time approaches \cite{li2023research,tian2022underwater}. Implementing a secondary decomposition assists in extracting high-level features and further denoising IMFs containing indistinguishable noise \cite{qiu2018ensemble}. VMD decomposes signals denoised by wavelet thresholding further \cite{liu2024application}. Then, the IMFs of high mutual information are selected for the following recognition tasks. Adaptive chirp mode decomposition, an advanced extension of EMD, is recently explored in \cite{li2023novel}.

\subsection{Variational mode decomposition}
\subsubsection{Theoretical development}
VMD can decompose the non-stationary signals into several sub-series called modes \cite{dragomiretskiy2013variational}. The VMD can be considered as the following problem 
 \begin{equation}
 \small
 	\label{eq: vmd1}
 	min \left\{m_{k}\right\},\left\{w_{k}\right\}\Biggl\{\sum_{k=1}^{K}||\delta_{t}\left[(\delta(t)+\frac{k}{\pi t})\times m_{j}(t)\right]e^{k\omega_{k}t}||^{2}_{2}\Biggr\}
 \end{equation}
with the constraints as
\begin{equation}
	\sum_{k=1}^{K}m_{k}=x(t),
\end{equation}
where $m_{k}$ is mode $k$, $\omega_{k}$ is $m_{k}$'s central frequency, $K$ is the number of modes, $x(t)$ represents the input time series. The problem shown in Equation \ref{eq: vmd1} is transformed into Equation \ref{eq:6} when introducing the $L_{2}$ penalty and Lagrange multiplier
\begin{multline}
	\label{eq:6}
 \tiny
	L(\left\{m_{k}\right\},\left\{w_{k}\right\},\lambda)=\alpha \Biggl\{\sum_{k=1}^{K}||\delta_{t}\left[(\delta(t)+\frac{k}{\pi t})\times m_{j}(t)\right] \\
  e^{k\omega_{k}t}||^{2}_{2}+ ||x(t)-\sum_{k=1}^{K}m_{k}|| \bigg \langle\lambda(t),x(t)- \sum_{k=1}^{K}m_{k} \rangle\Biggr\}.	
\end{multline}

The alternating direction method of multipliers (ADMM) algorithm is utilized to solve the above problem in VMD. Then, the modes $m_{k}$ and $\omega_{k}$ are obtained during the shifting process. According to the ADMM algorithm, the $m_{k}$ and $\omega_{k}$ can be computed from the following equations,
\begin{equation}
	\hat{m}_{k}^{n+1}=\frac{\hat{y}(\omega)-\sum_{i\neq k}\hat{m}_{k}(\omega)+\frac{\hat{\lambda}(\omega)}{2}}{1+2\alpha(\omega-\omega_{k})^{2}}
\end{equation}
\begin{equation}
	\hat{\omega}_{k}^{n+1}=\frac{\int_{0}^{\infty}\omega|\hat{m}_{j}(\omega)|^{2}d\omega}{\int_{0}^{\infty}|\hat{m}_{j}(\omega)|^{2}d\omega},
\end{equation}
where $n$ represents the number of iterations, $\hat{y}(\omega)$, $\hat{m}_{k}(\omega)$, $\hat{\lambda}(\omega)$ and $	\hat{m}_{k}^{n+1} $ represent the Fourier transform of $x(t)$, $m_{j}(t)$, $\lambda(t)$ and $m_{k}^{n+1} $, respectively.	

\subsubsection{VMD-based UAS denoising}

To overcome the theoretical limitations of EMD, \cite{dragomiretskiy2013variational} propose the variational mode decomposition (VMD) algorithm with solid theoretical development. VMD has successfully handled noisy UAS \cite{yang2020denoising,yang2023underwater,hu2019denoising}. For instance, \cite{yang2020denoising} employ VMD to decompose the input signal. Then, the authors apply the Savitzky-Golay filter and Lift wavelet threshold (LWTD) algorithms to denoise low-frequency and high-frequency components, respectively. Finally, all components are aggregated for reconstruction. Another principle of denoising IMFs is to classify noise-dominated and signal-dominated IMFs. Different denoising algorithms can be applied to noise-dominated and signal-dominated IMFs \cite{yang2021denoising}. For instance, wavelet-thresholding algorithm and Savitzky-Golay filtering are employed to denoise noise-dominated and signal-dominated IMFs, respectively \cite{yang2021denoising}. Their results demonstrate the superiority of VMD over EMD for UAS denoising tasks.

Unlike the above methods dividing IMFs into two groups (clean and noise, low-frequency and high-frequency), some studies divide IMFS into three groups, pure, mixed, and noisy signals for fine-gained denoising \cite{li2024new}. 

\subsection{Other decompositions}

The improved symplectic geometry modal decomposition generates IMFs in \cite{xing2024novel}. Unlike most literature, which utilizes some criterion to group IMFs into clean signal and noisy parts, spectral clustering is employed to cluster IMFs into mixed and noise clusters. Finally, wavelet thresholding techniques filter out noise in mixed clusters. The authors employ intrinsic time-scale decomposition and correlation coefficients to denoise UAS data \cite{li2022feature}.

\subsection{Thresholding}
Thresholding is a fundamental step in the decomposition-based denoising framework. It eliminates noisy information from all decomposed components based on a predefined threshold \cite{li2024new}. This technique comprises two main stages: threshold determination and the thresholding function application. The first stage, threshold determination, primarily involves calculating threshold values using an appropriate criterion. The second stage, the thresholding function, is concerned with removing noise components while preserving the significant signal elements according to the established threshold.

\subsubsection{Threshold determination}
When decomposing signals, distinguishing between meaningful components and noise is crucial. Noise components typically have little to no informational overlap with the original signal. The first stage of thresholding is to determine the threshold value. A good threshold value should assist in retaining signal-dominated information and eliminating noise as much as possible. Researchers have utilized a variety of criteria to compute the threshold value. For instance, correlation coefficients between IMFs and original signals are employed to determine the threshold \cite{yan2019mems}. Signal-dominated components should show a much higher correlation than noise-dominated components. However, correlation coefficients cannot measure the non-linear dependency between decomposed components and original UAS, which is essential in complex signal environments.

In addition to linear criterion, the entropy is an essential indicator to reflect information in each IMF \cite{li2018noise}. Hence, the literature has explored various entropy-based metrics to compute the threshold, such as permutation entropy (PE) \cite{li2018new,zhang2024new,li2019fusion,li2021double,xie2021optimized}, amplitude-aware PE \cite{li2018noise,li2024new}, dispersion entropy (DE) \cite{li2019denoising,li2020new,li2020novel}, fluctuation-based DE \cite{yang2023underwater,li2023hierarchical}, slope entropy \cite{li2021double}, weighted PE\cite{li2019novel}, neural network estimation time entropy \cite{li2022research}.

Mutual information quantifies the amount of information obtained about one random variable through another random variable. In the context of signals, it measures how much information the presence of one signal can tell about another signal. This is particularly useful when determining how much of one signal (such as the original) is present in another (like a decomposed signal component). Mutual information \cite{li2018new}.

In addition to the above thresholding technique based on single-threshold, dual thresholds are researched in the literature \cite{li2019denoising}. For instance, an interval thresholding is employed in \cite{li2019denoising}.

\subsubsection{Thresholding function}

Thresholding functions aim at eliminating noisy components while retaining strong signals. Table \ref{tab:Summary of representative thresholding functions.} summarizes basic thresholding functions and baselines for advanced thresholding functions in the literature \cite{5972413}. For instance, in \cite{wang2020novel}, a new adaptive thresholding function considering the continuity of input-output curves, and is defined as:
\begin{equation}
\footnotesize
\hat{w}_{j,k} = \begin{cases} 
  \text{sgn}(w_{j,k}) \left( |w_{j,k}| - \left| w_{j,k} \right|^\eta (\lambda_{j} - |w_{j,k}|) * \lambda_j \right), & |w_{j,k}| \geq \lambda_j \\
  0, & |w_{j,k}| < \lambda_j. 
\end{cases}
\end{equation}
Although advanced adaptive or semi-soft thresholding functions have been proposed and demonstrated outstanding performance, they often necessitate the optimization of additional hyper-parameters. This requirement can complicate their practical implementation and demand extensive computational resources or domain expertise to achieve optimal results. Such complexities can be a barrier, especially in applications with critical real-time processing or limited computational resources. Moreover, tuning these hyper-parameters can be sensitive to the specific characteristics of the data, making these methods less robust across diverse datasets unless carefully adjusted.
\begin{table}[htbp]
  \centering
  \caption{Summary of conventional thresholding functions.}
  \resizebox{.5\textwidth}{!}{
    \begin{tabular}{cccp{5cm}}
    \toprule
    Technique     & Formula    & Characteristic \\
    \midrule
    Hard thresholding \cite{5972413}&$f(x) = \begin{cases} 
  x_i, & |x| > \lambda \\
  0, & |x| \leq 0 
\end{cases}$&\begin{tabular}{@{}c@{}}Simplicity and efficiency \\ Non-Adaptivity \end{tabular}\\
    \hline
    Soft thresholding \cite{5972413}&$f(x) = \begin{cases} 
  \text{sgn}(|x| - \lambda), & |x| > \lambda \\
  0, & |x| \leq \lambda 
\end{cases}
$&\begin{tabular}{@{}c@{}}Smoothness in the Reconstructed Signal \\Over-Smoothing \end{tabular} \\
    \hline
    
    Semi-soft thresholding \cite{5972413}&$f(x) = \begin{cases} 
  \text{sgn}(x)(|x| - m\lambda), & |x| > \lambda \\
  0, & |x| \leq \lambda 
\end{cases}
$& \begin{tabular}{@{}c@{}}Flexible transfer between hard and soft schemes. \\ More hyper-parameters \end{tabular}\\
    \bottomrule
    \end{tabular}%
    }
  \label{tab:Summary of representative thresholding functions.}%
\end{table}%

\subsection{Hyper-parameters of signal decomposition}
A practical issue of the decomposition-based denoising framework is determining the hyper-parameters of decomposition algorithms \cite{liu2024application}. Signal decomposition algorithms share an essential hyper-parameter, the decomposition level. Smaller decomposition levels lead to significant mode mixing issues. However, high decomposition levels may generate components of fake frequencies and deteriorate the denoising performance. Meanwhile, each additional level of decomposition increases the computational burden. 

Researchers attempted to directly apply the decomposition level of EMD to VMD to retain the advantages of VMD while incorporating the adaptive capabilities of EMD \cite{liu2024application}. Spearman correlation coefficients are utilized as a threshold to determine whether the decomposed component is unsubtle \cite{yang2021denoising}. Measuring correlations between reconstructed and original UAS can also guide the selection of decomposition level \cite{yang2020underwater}. Evolutionary optimization successfully determines hyper-parameters of signal decomposition algorithms in the UAS denoising literature \cite{yan2019mems}. Researchers employ various evolutionary algorithms to search for threshold, decomposition, and other crucial parameters \cite{wang2020novel}.

\section{Deep learning}
Deep learning (DL) algorithms employ a deep neural network to reconstruct clean signals from noisy input \cite{koh2020underwater,testolin2019underwater,kumar2023self,xie2020improved}. Most literature has followed the framework of autoencoder for reconstruction \cite{song2023underwater}. Designing suitable architectures and novel loss functions to extract noise-resistant features efficiently is crucial \cite{mallik2022predicting,ji2024research,song2024underwater,ma2023underwater}. Table \ref{tab:Summary of representative DL-based UAS denoising studies} summarizes representative DL-based UAS denoising in recent years.
\begin{table}[htbp]
  \centering
  \caption{Summary of representative DL-based UAS denoising studies.}
  \resizebox{.5\textwidth}{!}{
    \begin{tabular}{ccccc}
    \toprule
    \multicolumn{1}{c}{Literature} & \multicolumn{1}{c}{Architecture} & \multicolumn{1}{c}{Input} & \multicolumn{1}{c}{Training logic} & \multicolumn{1}{c}{Hyper-parameters tuning} \\
    \midrule
    \cite{wang2020novel}     & MLP and CNN     & Original UAS  & Unsupervised reconstruction   & Manually \\
    \cite{tian2021deep}     & CNN     & Original UAS   & Supervised recognition    & Manually \\
    \cite{song2023novel}     & CNN and RNN    &Original UAS    & Unsupervised reconstruction     & Manually  \\
    \cite{zhou2023attention}     & RNN and Transformer    &STFT   & \begin{tabular}{@{}c@{}}Unsupervised reconstruction \\Supervised recognition \end{tabular}    & Manually  \\
    \cite{dong2022bidirectional}     & MLP    &Mel-frequency cepstral coefficients    & Unsupervised reconstruction     & Manually  \\
    \cite{zhou2021generative}     & CNN and GAN    &Original signal    & Unsupervised reconstruction     & Manually  \\
    \cite{ashraf2022ambient,ashraf2021underwater}     & GAN    &STFT    & Unsupervised reconstruction     & Manually  \\
    \cite{zhou2023dbsa}     & Attention    &STFT    & Unsupervised reconstruction     & Manually  \\
    \cite{zhou2023novelnoise}     & Attention    &STFT    & Unsupervised reconstruction     & Manually  \\
    \cite{song2022method}     & Attention    &Original UAS    & Unsupervised    & Manually  \\
    \cite{yang2024underwater}     & Attention    &Mel spectrum    & Recognition    & Manually  \\
    \cite{wang2024self}     & Attention    &Time-frequency spectrum    & Self-supervised   & Manually  \\
     \cite{xu2023underwater}     & CNN and attention    &Mel spectrum    & Supervised   & Manually  \\
      \cite{zhu2023underwater}     & CNN   &FFT    & Supervised   & Manually  \\
    \bottomrule
    \end{tabular}%
    }
  \label{tab:Summary of representative DL-based UAS denoising studies}%
\end{table}%

\subsection{DL-based UAS denoising methodology}
\begin{figure}[htbp]
	\centering
	\includegraphics[width=1\linewidth]{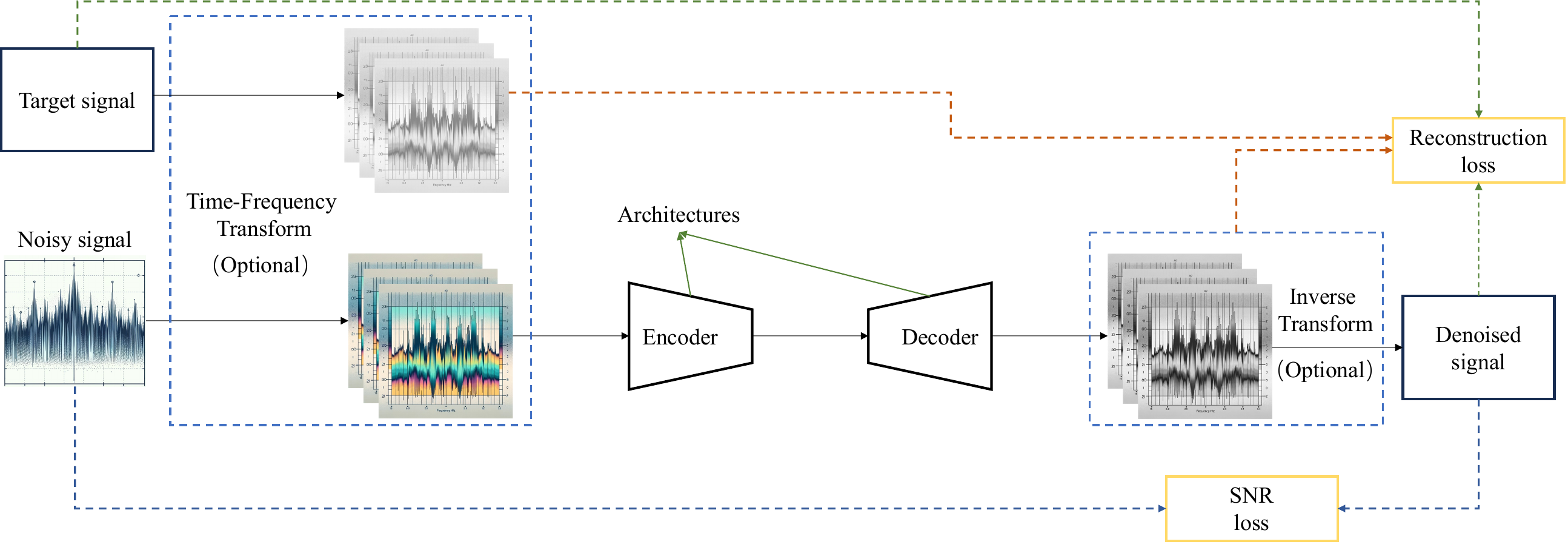}
	\caption{Framework of DL-based Autoencoder UAS denoising algorithms.}
	\label{fig:Framework of DL-based Autoencoder UAS denoising algorithms}
\end{figure}

Unlike traditional denoising algorithms and signal decomposition techniques, neural networks operate without preset assumptions about the noise characteristics \cite{ma2023underwater}. Various deep learning architectures have demonstrated efficacy in UAS denoising, including convolutional neural networks (CNNs) \cite{zhou2020denoising, sun2022underwater, tian2021deep, wang2014robust, chu2023deep, lyu2024light}, recurrent neural networks (RNNs) \cite{song2023novel}, and attention-based neural networks \cite{zhou2023attention}. The reconstruction-based DL denoising algorithms pipeline is visualized in Figure \ref{fig:Framework of DL-based Autoencoder UAS denoising algorithms}. Time-frequency transformation is optional because the DL model can directly process the original UAS. This framework trains a denoising DL model based on reconstruction loss and SNR-related loss. Reconstruction loss can be computed based on the spectrum when any Time-Frequency transform is adopted.

An ideal binary mask (IBM) is initially estimated using features derived from clean and noisy signals, followed by training a deep multilayer perceptron (MLP) to predict the IBM for effective denoising \cite{cao2023underwater}. To reconstruct the noisy input, a stacked convolutional sparse denoising autoencoder is employed, leveraging sparse representations \cite{wang2020stacked}. Furthermore, a Multiscale Residual Unit (MSRU) incorporating various convolutional kernels has been proposed to extract robust noise-resistant features \cite{tian2021deep}. Additionally, CNN features can be enhanced through a dual-path recurrent neural network, significantly improving the denoising performance for UAS \cite{song2023novel}. Considering the high dimensionality of the original time series, Mel-frequency cepstral coefficients (MFCCs) are extracted as training samples from both the original and denoised UAS using CEEMDAN \cite{dong2022bidirectional}.

Researchers have investigated the denoising capabilities of Generative adversarial networks (GANs) for UAS \cite{zhou2021generative, ashraf2021underwater}. Specifically, the GAN algorithm has been employed to mitigate underwater ambient noise \cite{ashraf2021underwater}. Initially, the short-time Fourier transform (STFT) is applied, using magnitude and phase features as inputs for the GAN. Clean signals are then reconstructed from the GAN's output using the inverse STFT (ISTFT). In another approach, a GAN is utilized to generate clean signals, with the discriminator designed to distinguish between real noisy signals and the combination of clean signals with ambient noise \cite{ashraf2022ambient}. This denoising model incorporates a 1D convolutional layer for feature extraction. Experimental results indicate that the GAN model surpasses both EMD and wavelet methods in performance.

Recently, the attention mechanism has been integrated into denoising UAS. \cite{zhou2023dbsa} developed a dual-branch attention-based neural network to reconstruct clean signals from noisy complex spectra. Comparative studies have demonstrated that this deep learning approach surpasses traditional Wavelet-based and EMD-based denoising algorithms. Additionally, \cite{zhou2023novelnoise} proposed a deep learning model incorporating Residual (Res) blocks and attention modules to effectively separate a noisy waveform into noise and a denoised waveform. Furthermore, a Transformer model, trained to maximize the SNR, has been utilized for acoustic signal denoising \cite{song2022method}. Another innovative deep learning model, featuring channel, frequency, and time attention modules, has been introduced to extract robust noise-resistant features across multiple domains \cite{yang2024underwater}.

Another significant challenge involves the disparate optimization objectives between denoising and recognition tasks in UAS processing \cite{zhou2023novel}. To address this, a joint training framework utilizing a modified Transformer neural network has been proposed, which successfully achieves both denoising and recognition \cite{zhou2023novel}. The loss function in this framework is dual-part; it includes a denoising component where the mean squared error between the noisy and the clean signal is minimized.

A self-supervised, dual-channel self-attention encoder has been proposed to learn robust, noise-resistant features of UAS \cite{wang2024self}. This self-supervised learning approach compels the UAS model to identify and retain the most informative and stable patterns for succeeding in the pretext task. Additionally, this method inherently encourages the model to learn features invariant to minor perturbations or variations (i.e., noise) in the input data, focusing on attributes crucial for distinguishing between fundamentally different classes or scenarios. Furthermore, data augmentation has been demonstrated to enhance the accuracy and noise robustness of the UAS model \cite{li2023data,xu2023underwater}.

Diversity in architectures is crucial for deep learning-based approaches in UAS processing. For example, multiple classifiers are constructed to handle different types of noise \cite{zhu2023underwater}. The pivotal principle in designing deep learning models for UAS-related tasks lies in extracting robust and noise-resistant features. A diverse set of feature extractors facilitates the extraction of multi-scale features, ultimately enhancing noise resistance. The literature demonstrates using convolutional filters with diverse kernel sizes to capture these multi-scale features \cite{zhou2023attention, li2024deep, li2021underwater, tian2023joint}. Different convolutional kernels aid in automatically learning features across various frequencies. Moreover, \cite{tian2023joint} propose a parallel architecture to jointly learn from UAS data, optimizing feature extraction for enhanced performance.

Designing suitable loss functions is essential for deep-learning denoising methods. Generally, DL-based UAS denoising algorithms employ reconstruction-based loss \cite{xu2024cross,zhou2020denoising,dong2022bidirectional}. They train a DL model to reconstruct clean signal-dominated components from noisy UAS \cite{zhou2023dbsa}. For recognition-based DL models, supervised classification losses are employed \cite{tian2021deep}. Besides reconstruction-based and recognition-based terms, other terms assisting in enhancing noise-resistance representations and generalizations are designed and employed \cite{zhou2020denoising}. For instance, the distance between learned features and feature centroid is minimized to enhance the noise-resistance of features \cite{li2024deep}. Correspondingly, a passive attention loss is defined. Some studies train the neural network to maximize SNR \cite{zhou2023novel,song2023novel}.
\subsection{Input formulation}
Appropriate formulations of the input for UAS are critical for the performance of deep learning denoising methods \cite{sun2022underwater}. Although directly processing raw UAS data is a straightforward choice, a meaningful formulation significantly aids in training deep learning models. Typically, the UAS community utilizes time-frequency transformations of the raw signal, such as the STFT \cite{yang2022open}, Mel-Spectrum \cite{yang2024underwater}, Bark spectrum \cite{xu2023underwater}, and hand-crafted features \cite{jiang2021interpretable, wang2023hidr}. Experimental studies have demonstrated that features like the magnitude STFT spectrum, complex-valued STFT spectrum, and log-mel spectrum notably enhance the performance of deep CNNs in underwater recognition tasks \cite{sun2022underwater}. Additionally, Mel-frequency cepstral coefficients are employed as inputs for a deep CNN to improve recognition accuracy further \cite{lyu2024light}.

\subsection{Data augmentation}
Data augmentation is a critical technique in improving the noise resistance of deep learning models and achieved significant success in sequential tasks \cite{li2022sample,li2023data}. Data augmentation artificially expands the training dataset by creating modified versions of the existing UAS. These modifications might include adding noise, cropping, or changing lighting conditions. This variety assists in training denoising and recognition models based on diverse samples. By training on a more diverse data set, data augmentation acts as a form of regularization. It effectively prevents the model from memorizing the training UAS (overfitting), encouraging more robust generalization abilities. 

UAS denoising models are usually developed based on the time-frequency transformation of the original UAS \cite{zhou2023attention}. Data augmentations can be directly applied to the original UAS in the time domain by adding noise \cite{zhou2023attention}. Adding noise is naturally advantageous for underwater tasks due to the noisy and complex characteristics of the underwater environment \cite{li2023data}. After transforming the original UAS into time-frequency representations, masking is a common and straightforward augmentation strategy \cite{xu2023underwater}. For instance, time-masking and frequency-masking are implemented on representations obtained from Mel filter bank \cite{zhou2023attention}.

Conventional data augmentation strategies in UAS processing involve training denoising or recognition models using both raw and augmented samples. However, the direct contribution of augmented samples to training losses may lead to performance degradation \cite{xu2023underwater}. To address this issue, a smoothness-inducing regularization technique has been proposed to minimize the distance between representations of raw and augmented samples, thereby improving the consistency and effectiveness of the training process \cite{xu2023underwater}. Additionally, a local masking and replicating technique has been developed, which randomly selects two samples, applies local masking, and mixes them to create a new augmented sample \cite{xu2023underwater}. Experimental results indicate that these proposed augmentation techniques outperform GAN-based algorithms in terms of enhancing model robustness and recognition accuracy \cite{gao2020recognition, yang2020gan}.

%\subsection{Loss function}

\section{Other methods}
In addition to the above UAS denoising techniques, the literature has explored other techniques, such as dictionary learning \cite{xing2021sparse} and Least mean squares (LMS) denoising \cite{yang2022denoising}. The study has demonstrated that the LMS denoising algorithm outperforms EMD and VMD \cite{yang2022denoising}.

\section{Evaluation metrics}
Performance evaluation is crucial for assessing the effectiveness of UAS denoising techniques \cite{xu2024cross}. Various metrics are utilized in the literature to evaluate UAS denoising performance. Predominantly, these metrics are based on the signal-to-noise ratio (SNR), which quantifies the desired signal level relative to the background noise. Additionally, some studies employ recognition accuracy as a direct measure of performance, primarily when denoising is intended to improve the accuracy of subsequent recognition tasks. This section summarizes the evaluation metrics commonly used in UAS denoising research.
\subsection{Signal quality metrics}
\subsubsection{Signal-to-noise ratio}
The signal-to-noise ratio (SNR) quantifies the proportion of signal power to noise power. Higher SNR values indicate lower noise content in the signal, whereas lower SNR values suggest higher noise content. The SNR is defined as:

\begin{equation}
SNR = 10 \log_{10} \left( \frac{P_{signal}}{P_{noise}} \right),
\end{equation}
where $P_{signal}$ and $P_{noise}$ represent the power of pure UAS and noisy signals, respectively. SNR is a necessary and popular evaluation metric for UAS denoising \cite{yang2020denoising}.

\subsubsection{Peak signal-to-noise ratio}
The peak signal-to-noise ratio (PSNR) is a significant metric for evaluating UAS denoising quality. For the UAS, PSNR measures the ratio of the maximum possible power of a signal to the power of corrupting noise that affects the fidelity of its representation \cite{ma2023underwater}. The PSNR can be computed by
\begin{equation}
\text{PSNR} = 10 \log_{10} \left( \frac{\text{MAX}_I^2}{\text{MSE}} \right),
\end{equation}
where MSE is the mean squared error between the original and the denoised signal. A higher PSNR value indicates that the denoised signal is higher quality than the noise level. The denoising process effectively reduces the noise while preserving the integrity and strength of the original signal.

\subsubsection{Signal-to-distortion ratio}

The signal-to-distortion ratio (SDR) specifically focuses on the distortion between the original signal and the estimated signal \cite{zhou2023dbsa}. The SDR is defined as,

\begin{equation}
SDR = 10 \log_{10} \left( \frac{P_{signal}}{P_{distortion}} \right),
\end{equation}

\subsubsection{Signal-to-distortion ratio improvement}
The Signal-to-Distortion Ratio Improvement (SDRi) measures the improvement in SDR due to some processing or alteration of a signal. The SDRi is calculated by comparing the SDR before and after the processing \cite{zhou2023dbsa}. 

\begin{equation}
SDRi = SDR_{\text{after}} - SDR_{\text{before}}
\end{equation}

\subsubsection{Scale-invariant signal-to-noise ratio improvement}

The Scale-Invariant SNR Improvement (SI-SNRi) is a measure often used in audio and speech processing to evaluate the effectiveness of enhancement algorithms, particularly when the absolute scale of the signal may not be consistent or important. This metric adjusts for scaling differences between the processed and original signals, providing a more robust comparison.

\begin{equation}
\text{SI-SNRi} = \text{SI-SNR}_{\text{after}} - \text{SI-SNR}_{\text{before}}
\end{equation}

The scale-invariant SNR is calculated differently from the traditional SNR to account for scaling factors between the target and estimated signals. It involves normalizing the signal relative to a reference before computing the power ratio:
\begin{equation}
\text{SI-SNR} = 10 \log_{10} \left(\frac{\|\alpha \cdot x(n)\|^2}{\|\hat{x}(n) - \alpha \cdot x(n)\|^2}\right),
\end{equation}

 where $\alpha = \frac{\langle \text{estimate}, \text{target} \rangle}{\|\text{target}\|^2}$ scales the target signal to best fit the estimate in a least-squares sense. This normalization allows the SI-SNR to be independent of the signal's scale, focusing solely on the noise and distortion relative to the target's shape and structure \cite{zhou2023dbsa}.

\subsubsection{Segment signal-to-noise ratio}
The Segment SNR (SSNR) averages the SNR values computed for each segment, giving a more detailed measure of signal quality across different parts of the signal, which is especially useful in cases where signal characteristics vary over time \cite{zhou2023dbsa}. The SSNR can be computed by
\begin{equation}
SSNR = \frac{10}{M} \sum_{m=1}^M \log_{10} \left( \frac{P^{m}_{signal}}{P^{m}_{noise}} \right),
\end{equation}
where $P^{m}_{signal}$ and $P^{m}_{noise}$ represent the power of the $m^{th}$ segment of the signal and noise, respectively.

\subsection{Reconstruction error}
Reconstruction error measures the deviations between the pure signal and noise reduction signal. An outstanding denoising algorithm should precisely reconstruct the pure signal and achieve a small reconstruction error. There are various reconstruction errors in the literature \cite{yang2020denoising}. Table \ref{tab: Forecasting metrics.} summarizes popular reconstruction errors employed in the UAS denoising literature.

\begin{table}[htbp]
\centering
  \caption{Reconstruction error in the UAS denoising literature.}
  \label{tab: Forecasting metrics.}
  \resizebox{.5\textwidth}{!}{
\begin{tabular}{ccc}
\hline
Metric& Definition&Literature \\
\hline
Mean square error (MSE)&$\frac{1}{L}\sum_{n=1}^{L}(\hat{x_{n}}-x_{n})^2$&\cite{yang2020new,yang2023underwater,yang2024underwater}\\
Root Mean square error (MSE)&$\sqrt{\frac{1}{L}\sum_{n=1}^{L}(\hat{x_{n}}-x_{n})^2}$&\cite{al2017improved,li2018noise,yang2020denoising,yang2020new,li2020novel}\\
Mean absolute error (MAE)&$\frac{1}{L}\sum_{j=1}^{L}|\hat{x_{n}}-x_{n}|$&\cite{baskar2015study}\\
\hline
\end{tabular}}
\end{table}

%\subsection{Divergence}
%\cite{Kullback-Leibler divergence, Jensen-Shannon divergence}

%\cite{zhou2021generative}

%\subsection{Recognition accuracy}

%The ultimate objective of denoising is to enhance the precision of recognition tasks. To this end, certain studies utilize recognition accuracy as a metric to evaluate denoising efficacy \cite{dong2022bidirectional}. However, using recognition accuracy to assess denoising performance directly presents challenges. This is because highly effective modules at classification may compensate for deficiencies in the denoising process, thus obscuring the actual performance of the denoising efforts.

\section{Experimental dataset}
The literature on UAS denoising algorithms has been evaluated using a variety of datasets due to the challenges inherent in the underwater environment and the difficulties associated with real-world data collection. Consequently, many studies have utilized synthetic data and artificial noise to test their algorithms. On the other hand, some studies have conducted experiments with real-world UAS datasets. The table below provides a summary of popular datasets used in UAS denoising research.
\subsection{Synthetic data}
Due to the limited availability of UAS datasets, researchers have purposefully simulated synthetic pure and noise signals to test denoising algorithms. Table \ref{tab:Summary of synthetic pure signals.} summarizes synthetic pure signals are popular in the UAS denoising literature. The literature may generate different pure signals with different initial states. Then, various synthetic noise signals are added to pure signals to simulate the noisy UAS. According to Table \ref{tab:Summary of synthetic pure signals.}, Lorenz signal is the most popular, whereas Ikeda and Mackey Glass signals are much less popular.

\begin{table}[htbp]
  \centering
  \caption{Summary of synthetic pure signals.}
  \resizebox{.5\textwidth}{!}{
    \begin{tabular}{cccp{5cm}}
    \toprule
    Signal    & Formula    & Literature \\
    \midrule
    Lorenz signal&$\begin{cases}
\dot{x} = -r(x - y) \\
\dot{y} = -xz + gy - y \\
\dot{z} = xy - kz
\end{cases}
$&\begin{tabular}{@{}c@{}}\cite{li2023research,li2018new,li2018noise,yang2021denoising,li2022feature,li2022research}\\ \cite{li2024new,yang2023underwater,li2024noise,yang2023dual} \end{tabular}\\
    \hline
    Chen signal &$\begin{cases}
\dot{x} = -r(x - y) \\
\dot{y} = (k - r)x - xz + ky \\
\dot{z} = xy - gz
\end{cases}

$&\cite{li2024new,yang2021denoising,yang2023underwater,li2024noise,li2022feature} \\
    \hline
    
    Rossler signal &$\begin{cases}
\dot{x} = -y - z \\
\dot{y} = x + ry \\
\dot{z} = g + z(x - k)
\end{cases}$& \cite{li2024new,yang2023underwater,li2024noise,yang2023dual}\\
\hline
Duffing signal &$\begin{cases}
\dot{x} = y \\
\dot{y} = -\rho y + \alpha x (1 - x^2) + \mu \cos(\omega t) \\
\dot{z} = \delta
\end{cases}$& \cite{li2024new,yang2023dual,li2022feature}\\
\hline
Rossler signal &$\begin{cases}
\dot{x} = -y - z \\
\dot{y} = x + ry \\
\dot{z} = g + z(x - k)
\end{cases}$& \cite{li2024new,yang2023underwater,li2024noise,yang2023dual}\\
\hline
Mackey Glass signal &$\begin{cases} 
    \frac{dx(t)}{dt} = \beta \frac{x(t-\tau)}{1 + x(t-\tau)^n} - \gamma x(t) & \text{for } x(t-\tau) > 0, \\
    z = \theta & \text{otherwise.} % Assuming 'z' and '\theta' are placeholders
\end{cases}$& \cite{yang2023dual}\\

\hline
Ikeda signal &$\begin{aligned}
x_{n+1} &= 1 + u \left( x_n \cos(t_n) - y_n \sin(t_n) \right), \\
y_{n+1} &= u \left( x_n \sin(t_n) + y_n \cos(t_n) \right), \\
t_n &= \kappa - \frac{\alpha}{1 + x_n^2 + y_n^2},
\end{aligned}$& \cite{yang2023dual}\\

    \bottomrule
    \end{tabular}%
    }
  \label{tab:Summary of synthetic pure signals.}%
\end{table}%

%The literature also utilized synthetic noise to investigate the denoising ability of the proposed algorithm \cite{zhou2023novel,yang2020denoising}. For instance, \cite{yang2020denoising} add Gaussian white noise of -5dB, 0dB, 5dB, 10dB, and 15dB to pure signals. \cite{zhou2023dbsa} utilize three types of underwater ambient noise, including rain, flow, and wind noise, to simulate complex underwater environments. The authors synthesize noisy samples with low SNR from -15 dB to -5 dB. Furthermore, the authors also investigate the superposition of three kinds of ambient noise. 

\subsection{Real-world data}
\begin{table}[htbp]
  \centering
  \caption{Summary of real-world dataset.}
  \resizebox{.5\textwidth}{!}{
    \begin{tabular}{cccp{5cm}}
    \toprule
    Dataset    & Access   & Literature \\
    \midrule
    ShipsEar& https://underwaternoise.atlanttic.uvigo.es/ \cite{santos2016shipsear}&\begin{tabular}{@{}c@{}}\cite{zhou2023attention,xu2023underwater,yang2024underwater,chen2024ship}\\ \cite{zhou2023dbsa,song2023novel,dong2022bidirectional,ashraf2022ambient,zhou2023novelnoise,wang2024self,li2023data} \end{tabular}\\
    \hline
    DeepShip &https://github.com/irfankamboh/DeepShip \cite{irfan2021deepship}&\begin{tabular}{@{}c@{}}\cite{zhu2023underwater,chen2024ship}\\ \cite{zhou2023attention,xu2023underwater,yang2024underwater,chen2024ship} \end{tabular} \\
    \hline
    Cooperative-project Dataset &Confidential& \cite{wang2024self}\\
    \hline
    Ocean Networks Canada &https://oceannetworks.ca& \cite{wang2024self}\\
    \hline
    Dataset of two type of vessels &Confidential& \cite{ren2019feature,xu2023underwater}\\
    \hline
    Ship radiated noise &https://aigei.com/& \cite{li2024new}\\
    \hline
    Department of the Interior of National Parks&https://www.nps.gov/glba/learn/nature/soundclips.htm&\cite{li2024noise}\\
\hline
National Park Administration&https://www.nps.gov/glba/learn/nature/soundclips.html& \cite{yang2021denoising}\\
    \bottomrule
    \end{tabular}%
    }
  \label{tab:Summary of real-world signals.}%
\end{table}%
In addition to the synthesized data discussed earlier, the literature also examines UAS denoising algorithms using real-world datasets. Table \ref{tab:Summary of real-world signals.} provides a summary of studies that utilize these real-world datasets. Unfortunately, most datasets gathered by respective authors are not publicly accessible. Among the publicly available datasets, ShipsEar and DeepShip are the most notable. The ShipsEar dataset comprises underwater acoustic recordings of ships and boats, featuring 90 recordings across 11 vessel types, totaling 6189 seconds of audio. In contrast, the DeepShip dataset includes 47 hours and 4 minutes of real-world underwater recordings, capturing 265 different ships categorized into four classes. These recordings were made throughout various seasons, featuring diverse sea states and noise levels. The DeepShip dataset contains nearly seven times more recordings and is approximately 25 times longer in total duration than the ShipsEar dataset. Usually, each record is segmented into small windows to train denoising algorithms \cite{dong2022bidirectional}.

\section{Applications}
UAS denoising is a necessary component for various underwater applications. This section presents some major roles of UAS denoising technologies in real-world applications.

1. \textbf{Maritime Navigation and Safety} 

Improved denoising techniques help in clearer detection of obstacles, other vessels, and navigational aids, reducing the risk of collisions and grounding in poor visibility conditions \cite{austin1994application,alcocer2006underwater}. 

2. \textbf{Submarine Communications }

In underwater environments, where radio waves cannot be effectively used, acoustic signals serve as the primary communication medium \cite{song2019editorial}. Denoising these signals ensures more reliable and clearer communications between submarines and surface vessels. 

3. \textbf{Marine Life Monitoring }

Acoustic signals are used to monitor the presence, movement, and behavior of marine species. Effective denoising is essential to accurately identify species from their sounds, which is crucial for ecological studies and conservation efforts \cite{rako2019underwater}. 

4. \textbf{Shoreline surveillance}

Shoreline surveillance refers to the monitoring and observation activities conducted along coastlines to ensure security, safety, and environmental integrity. This practice involves the use of various technologies and strategies to detect, track, and respond to activities and natural phenomena that occur near the shore \cite{domingos2022survey}.

\section{Open questions and future directions}
The UAS community has delved into advanced decomposition frameworks, thresholding techniques, and DL algorithms. Nevertheless, numerous unexplored avenues remain, warranting further extensive research and exploration.
\begin{itemize}
    \item[1] Signal decomposition is widely used to denoise the UAS, yet identifying the optimal decomposition level continues to pose a significant challenge. While the sensitivity analysis of decomposition levels across different algorithms has been explored, it remains largely cursory. Treating the decomposition level as a hyperparameter and applying hyperparameter optimization algorithms could be a promising strategy. However, this approach is often too time-consuming and impractical for underwater applications, which demand real-time processing capabilities.
    %the best singal decomposition
    \item[2] There is no consensus in the literature regarding the most effective signal decomposition algorithm for denoising the UAS. Benchmarking studies comparing different signal decomposition algorithms are notably absent. While some studies claim the superiority of specific algorithms based on outcomes from signal decomposition research, UAS present unique challenges that necessitate further specialized investigation.
    %standard framework
    \item[3] The literature employs different steps and parameters, such as normalization, sampling rate, and window length, to preprocess the UAS data before establishing the following denoising model. Differences in preprocessing lead to different conclusions and findings in terms of the performance of denoising and recognition models. A specific UAS denoising model may obtain outstanding performance on a small window and become much worse on long windows. Therefore, the standard framework to identify preprocessing schemes for any UAS dataset needs to be well-researched.
    %randomized learning
    \item[4] Although researchers have explored various advanced DL architectures, underwater applications require extremely high computing speed. When deploying the technology in underwater scenarios, the denoising model must adapt to new contexts quickly. Therefore, gradient-based DL models may fail to satisfy these requirements. However, deep randomized neural networks are suitable candidates due to their strong non-linear feature extraction ability and fast training speed \cite{malik2023random}.
    %automatic learning
    \item[5] Automated learning, encompassing the fine-tuning of hyperparameters and training of learning-based UAS denoising algorithms, is imperative. The varied applications of UAS denoising techniques necessitate an automated framework, enabling practitioners to seamlessly employ these methods across diverse applications. Regrettably, current research overlooks the critical need for and significance of automated learning. Manual selection of hyperparameters and training algorithms diminishes the flexibility and practicality of UAS denoising techniques. Therefore, developing a environment-agnostic automatic learning framework of the UAS denoising techniques is worthy to explore.
    %ensemble denoising
    \item[6] Advanced ensemble learning techniques have demonstrated their efficacy in enhancing model robustness through the creation of a diverse array of base models \cite{ganaie2022ensemble}. In the realm of UAS-related tasks, the extraction of noise-resistant features and the development of robust models are imperative. However, the existing literature largely neglects the significance of an ensemble UAS denoising framework. Such a framework holds the potential to mitigate the weaknesses inherent in individual UAS denoising models, thereby enhancing overall denoising performance. Given the diversity and complexity of underwater environments, a single UAS denoising technique may struggle to achieve optimal performance across all scenarios. Nevertheless, through ensemble learning, various base models can collectively address different types of noisy signals, thus culminating in improved accuracy. 

    \item[7] Efficient exploration of the marine environment and accurate detection of underwater events necessitate the deployment of diverse data collection systems, including UUV swarms and acoustic sensor networks. Each individual UUV and sensor unit possesses the capability to gather acoustic signals from distinct spatial coordinates. However, the integration of these disparate UUV systems and the development of a denoising algorithm based on the amalgamated data remains an uncharted territory. The potential schemes for dynamically fusing these data sources and denoising the resulting signal hold significant promise and merit further exploration.

    \item[8] Many UAS denoising algorithms operate within an offline framework, assuming all UAS data is available for model establishment. However, as UAS data is inherently sequential and underwater applications often occur in real-time scenarios, online processing is essential. The denoising step of UAS data typically precedes control or detection systems, which operate continuously in real-world applications. Thus, there's a need to extend decomposition-based and DL-based UAS denoising techniques to online variants. These online algorithms can incorporate online signal decomposition methods, real-time training algorithms for DL models, and online adaptation of DL architectures, among other considerations.

    \item[9] While many UAS denoising algorithms rely on unsupervised reconstruction, a few pioneering studies have delved into the potential of self-supervised learning for UAS denoising. However, a plethora of advanced self-supervised learning techniques remain unexplored and under-investigated. Crafting appropriate pretext tasks is crucial for the efficacy of self-supervised learning algorithms, especially given the unique challenges posed by UAS data. Moreover, there is ample opportunity to design pretext tasks tailored to UAS characteristics. Furthermore, integrating self-supervised learning with existing decomposition-based denoising algorithms holds promise. The development of a self-supervised decomposition-based denoising scheme represents a particularly promising avenue for future research.
    \end{itemize}

\section{Conclusion}

Underwater acoustic signals (UAS) are the most commonly collected data in underwater environments and play a pivotal role in a variety of applications. However, the inherent complexity of these environments poses significant challenges for the transmission, recognition, feature extraction, and interpretation of UAS. As a result, denoising UAS is a critical technological step essential for various applications. This review paper provides an overview of the developments in UAS denoising, from theoretical underpinnings to practical applications. Denoising involves the removal of extraneous noise and the extraction of signal-dominated information, which is then utilized for tasks such as target recognition. Traditional methods typically rely on signal processing and wavelet thresholding algorithms for noise removal. The majority of UAS denoising solutions employ signal decomposition techniques, which facilitate the separation of the complex original UAS into multiple modes. Subsequently, specific denoising algorithms are applied to each mode to eliminate noise. Finally, all denoised modes are recombined to produce the cleaned UAS.

Recently, the rapid advancement of deep learning (DL) algorithms has spurred the UAS community to develop sophisticated DL-based denoising methods. These algorithms are typically trained to reconstruct signal-dominated information and maximize the signal-to-noise ratio. Given that a DL model's loss function can comprise multiple terms, researchers have the ability to combine denoising and recognition objectives within a single loss function. Consequently, DL-based denoising is not only task-oriented but also highly flexible, adapting to specific application needs with greater efficacy.

%\begin{thebibliography}{1}
%\end{thebibliography}
\small
\bibliographystyle{IEEEtran}%{unsrt}
\bibliography{Output}
%Biography text here without a photo.

%\begin{IEEEbiography}[{\includegraphics[width=1in,height=1.25in,clip,keepaspectratio]{fig1.png}}]{IEEE Publications Technology Team}
%In this paragraph you can place your educational, professional background and research and other interests.\end{IEEEbiography}

\end{document}